\documentclass[%
reprint,
superscriptaddress,
footinbib,
 amsmath,amssymb,
 aps,
]{revtex4-2}

\usepackage{graphicx}
\usepackage{dcolumn}
\usepackage{bm}
\usepackage{lmodern}
\usepackage[T1]{fontenc}

\begin{document}

\preprint{APS/123-QED}

\title{Nodal-line driven anomalous susceptibility in ZrSiS}
\author{Bruno Gudac}
\affiliation{Department of Physics, Faculty of Science, University of Zagreb, 10000 Zagreb, Croatia}
\author{Markus Kriener}
\affiliation{RIKEN Center for Emergent Matter Science (CEMS), Hirosawa 2-1, Wako-shi, Saitama 351-0198, Japan}
\author{Yuriy V. Sharlai}
\affiliation{B. Verkin Institute for Low Temperature Physics $\&$ Engineering, Ukrainian Academy of Sciences, Kharkiv 61103, Ukraine}
\author{Mihovil Bosnar}
\affiliation{Division of Theoretical Physics, Ru\dj{}er Bo\v{s}kovi\'c Institute,  10000 Zagreb, Croatia}
\affiliation{Donostia International Physics Center, 20018 Donostia-San Sebastian (Gipuzkoa) Spain}
\author{Filip Orbani\'c}
\affiliation{Department of Physics, Faculty of Science, University of Zagreb, 10000 Zagreb, Croatia}
\author{Grigorii P. Mikitik}
\affiliation{B. Verkin Institute for Low Temperature Physics $\&$ Engineering, Ukrainian Academy of Sciences, Kharkiv 61103, Ukraine}
\author{Akio Kimura}
\affiliation{Graduate School of Advanced Science and Engineering, Hiroshima University, Higashi-Hiroshima 739-8526, Japan}
\affiliation{Graduate School of Science, Hiroshima University, Higashi-Hiroshima, Hiroshima 739-8526, Japan}
\author{Ivan Kokanovi\'c}
\affiliation{Department of Physics, Faculty of Science, University of Zagreb, 10000 Zagreb, Croatia}
\author{Mario Novak}
\email{mnovak@phy.hr}
\affiliation{Department of Physics, Faculty of Science, University of Zagreb, 10000 Zagreb, Croatia}
\date{\today}

\begin{abstract}
We demonstrate  a unique  approach to test the signature of the nodal-line  physics by  thermodynamic methods.      
By measuring  magnetic susceptibility in    ZrSiS we found  an intriguing  temperature-driven crossover from dia- to paramagnetic behavior.
 We show that the anomalous behavior represents a real thermodynamic signature of the underlying nodal-line physics through the means of chemical pressure (isovalent substitution of Zr for Hf), quantum oscillations, and theoretical modeling.
 The anomalous   part of the susceptibility is orbital by nature, and it arises due to the vicinity of  the Fermi level to a   degeneracy point created by   the crossing   of  two nodal lines. 
Furthermore, an unexpected Lifshitz topological transition at the degeneracy point is revealed by tuning the Fermi level. 
The present findings in ZrSiS give a new and attractive starting point for various nodal-line physics-related phenomena to be tested by thermodynamic methods in other related materials.

\end{abstract}
\maketitle
Dirac materials  are characterized by a linear dispersion of energy-momentum curves near the Fermi energy. 
Due to the linearity of the  dispersion,  quasi-particles generically exhibit low effective masses and high mobilities \cite{Geim2007}. 
Dirac states  have been realized and studied in plenty of  materials and new fascinating phenomena such as  Fermi arcs and  chiral anomalies have been revealed \cite{Armitage2018,Xiong2015}. 
Under high-magnetic fields,  these materials exhibit  remarkably 
different behavior than  conventional matter. 
The Landau levels become non-equidistant with a square root magnetic field dependence \cite{McClure1956}.
This feature is  often used to  verify  the Dirac nature of a material experimentally \cite{Plochocka2008, Martino2019}. 

The magnetic susceptibility, a low-field limit of the magnetic response function, in  Dirac and nodal-line systems, shows unique properties due to the existence of  band touching degeneracy points and/or lines \cite{Mikitik1989, Mikitik2016, Koshino2007, Koshino2016, Ominato2018}.
For example, in contrast to ordinary metals, in  Dirac materials  the orbital (Landau) susceptibility diverges as the Fermi energy approaches the  degeneracy point at low temperatures \cite{Mikitik1989, Mikitik2016, Koshino2007, Koshino2016}.
The orbital susceptibility originates from  the dynamics of the  Bloch electrons.  
It incorporates  interband magnetic field mixing, and it is summed over  Bloch states \cite{Ogata2015,Fukuyama1971}.
Although it possesses unique fingerprints, the  susceptibility of  novel  Dirac systems has not been widely studied  experimentally \cite{Raoux2014, Nair2018,Fuseya2015}.

\begin{figure*}   
\includegraphics[width=18cm]{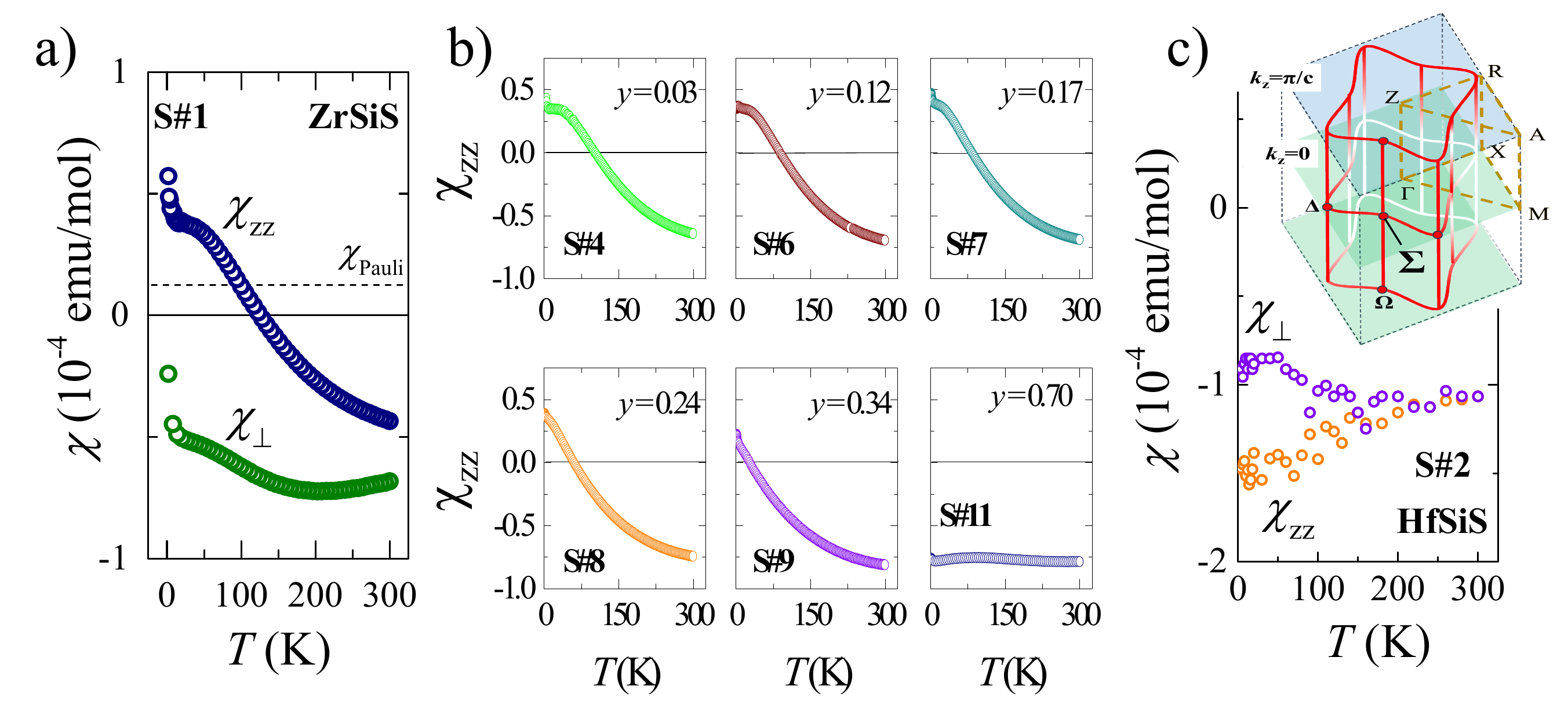}
\caption{\label{fig1} Temperature ($T$) dependent magnetic susceptibility  of Zr$_{1-y}$Hf$_y$SiS taken at 1T  for various Hf concentrations $y$.  a) Out-of-plane $\chi_{zz}$, along [001], and in-plane $\chi_{\perp}$ susceptibility of  ZrSiS. 
$\chi_{zz}$ has a strong anomalous $T$ dependence with unexpected change in the susceptibility from diamagnetic at high $T$ to paramagnetic at low $T$. 
By contrast $\chi_{\perp}$ exhibits a weak $T$ dependence with an impurity tail at low $T$.    $\chi_{Pauli}$ (dashed line) indicates the free-electron Pauli paramagnetic contribution as estimated from the specific heat measurements.    
b) $\chi_{zz}$ for different $y$: Upon increasing chemical pressure the sign change in $\chi_{zz}$ shifts toward lower $T$. For  $y>0.24$  the low-$T$ saturation of $\chi_{zz}$ vanishes. At high doping $y=0.7$, $\chi_{zz}$ flattens and remains diamagnetic in the whole  $T$ range examined.
 c) $\chi_{zz}$ and $\chi_{\perp}$ for HfSiS exhibit  a conventional, almost isotropic and  week $T$ dependence. (Inset) Schematics of the cage of  nodal lines  in the  1st Brillouin zone.  
The  intersecting nodal-line of interest  $\Sigma$ is  located along the $\Gamma$-$M$ direction.}
\end{figure*}

In this letter,  we present  a chemical pressure study of the magnetic susceptibility of the  nodal-line Dirac semimetal ZrSiS.
The susceptibility exhibits  an anomalous temperature dependence   pointing towards a unique  thermodynamic signature of  nodal-line physics. 
The  anomaly arises due to  the Fermi level being  located sufficiently close to the degeneracy point formed by  the crossing of  two nodal lines.

ZrSiS  crystallizes in a tetragonal square-net crystal structure with  nonsymmorphic P4/nmm (129) symmetry \cite{Tremel1987}.
It is one of the first reported topological nodal-line semimetals and has been the most extensively studied  among a range of different  nodal-line materials \cite{Xu2015,Schoop2016, Neupane2016, Hirayama2017,Pezzini2018,Shao2020,Gatti2020,Fu2019,Orbanic2021}.
In ZrSiS, there are  two types of  nodal lines.  
Topologically protected nodal lines, nonsymmorphic symmetry protection, are positioned with about 1eV below the Fermi level  and are not  significant for our experimental  findings \cite{Schoop2016, Neupane2016}. 
However, close to the Fermi level, another set of nodal lines is found. 
Those nodal lines form a cage-like structure, see the inset to  Fig.\ref{fig1}c. 
They  are not topologically protected, only C4v symmetry is present, and are thus   prone to the opening of a gap that is induced by the spin-orbit  (SO) interaction. 
Due to the relatively weak SO coupling, the resulting gap   is small \cite{Uykur2019}. 
 
Angle-resolved-photoemission spectroscopy (ARPES) and  high magnetic field studies  together  with   first principle calculations yield  a reasonably  well understood  Fermi surface (FS) \cite{Fu2019, Pezzini2018, Muller2020, Novak2019}.
The FS is wrapped around the cage-like nodal lines. It consists of a pair of  strongly anisotropic electron and hole pockets with open orbits along the $k_z$ direction \cite{Pezzini2018}. 
At the Fermi  level only the   Dirac-like  bands are present,  while the  so-called "trivial" bans are well away \cite{Schoop2016}.

Recent high-pressure studies  reported a change in the Berry phase of the quantum oscillations (QO),   indicating the possibility of  a topological phase transition (TPT).
In Ref.\cite{VanGennep2019} the authors  report the possibility  of  a TPT at pressures as low as 0.5 GPa due to a  phase change in  the QO frequency mode $F_{\delta2}$, this paper's notation. 
Additionally, the authors in  Ref.\cite{Gu2019}  report another TPT at around 7 GPa with a transition  detected in  the QO frequency mode  $F_{\alpha}$, again in the notation used here. 
In contrast,  in our  study, we observe a significantly  different qualitative  behavior,  even in the low-chemical pressure limit, i.e., at low Hf concentrations ($y$).
We have detected two possible chemical-pressure-driven Lifshitz transitions  in the low-frequency spectrum of the  QOs. 
The first transition is associated with the emergence  of a new frequency $F_{\psi}$  around $y=0.24$. 
The second  transition is observed  at around $y\approx0.34$ as indicated by  the disappearance of the pocket $F_{\delta1}$. 
The notable  difference between the results of  hydrostatic and chemical pressure could be traced back to a different compression of the
crystal lattice: In the case of chemical substitution of Zr
by Hf Vegard’s law is obeyed, i.e., the tetragonal lattice  ratio $c/a=2.273$ is practically constant across Zr$_{1-y}$Hf$_y$SiS, while that is not the case for the hydrostatic pressure where the  ratio is reduced by increasing pressure \cite{Tremel1987,Gu2019}.

Single crystals of Zr$_{1-y}$Hf$_y$SiS were grown by  chemical vapor transport starting  from a polycrystalline mixture of high-purity elements and iodine as a transport agent. 
With optimized growth conditions, well shaped prismatic crystals of several mm$^3$ weighing 20-30 mg that are ideal for magnetization studies were observed. 
To get information on the concentration and homogeneity of the doped  Hf atoms,  the  samples   were characterized  by employing an    Electron Probe Micro Analysis and x-ray diffraction (XRD).  
XRD patterns show a systematic shift of the  (002) peak upon increasing the Hf concentration without any traces of  segregation, indicating a homogeneous Zr-Hf solid solution.

Magnetic susceptibility ($\chi$) vs. temperature ($T$)   was measured using a Quantum Design MPMS SQUID magnetometer from 300 K to 1.8 K in the linear response  regime  at a field of 1T. 
The out-of-plane susceptibility  $\chi_{zz}$ was measured along the [001] direction, and  the in-plane  susceptibility   $\chi_{\perp}$ was measured perpendicular to [001].
There was no  difference in the sample response  between field cooled (FC) and zero field cooled  (ZFC) measurements.     The de Haas–van Alphen   oscillations were measured in [001] direction in the field range -7 T to 7 T at several different temperatures in order to estimate the quasi-particle  effective mass. 
The QOs were not significantly damped upon increasing the Hf concentration  indicating a small level of atomic substitution-induced disorder.

Figure \ref{fig1} summarizes magnetic susceptibility data of Zr$_{1-y}$Hf$_y$SiS.
For $y=0$, there is  a strong  anisotropy between $\chi_{zz}$ and $\chi_{\perp}$, cf. Fig.\ref{fig1}a. 
 Moreover, $\chi_{zz}$ has an unconventional step-like temperature dependence with  a  $T$-driven transition from a dia- to paramagnetic state and saturation   towards the  lowest $T$, whereas  $\chi_{\perp}$ is  relatively weakly  $T$-dependent. 
 The  low-$T$ paramagnetism of ZrSiS is  much stronger than  expected for free-electron Pauli susceptibility term  $\chi_{Pauli}$ as derived from the specific heat  density of states.  
The small upturns at the lowest temperatures (below $\sim$10K) originate  from paramagnetic impurities  of 0.01$\%$. 
 On the other  hand, the end compound HfSiS  (Fig.\ref{fig1}c) exhibits more  isotropic and $T$-independent susceptibility, which is more in  line with expectations for conventional materials.
To get better insight into the  unconventional behavior of  $\chi_{zz}(T)$ observed for $y=0$ we introduce  positive chemical pressure by  replacing Zr with isovalent Hf. Respective susceptibility measurements for selected $y$ are shown in Fig.\ref{fig1}b. 
Upon increasing the  Hf content, the step-like feature indicative of the transition from  dia- to paramagnetic behavior shifts to lower-$T$. 
For $y=0.34$ the susceptibility starts to deviate from the step-like behavior,   although the dia- to paramagnetic  transition is still present. 
However, it is significantly weaker and no longer saturates  towards low $T$, indicating that the effect responsible for this unconventional behavior is suppressed by  chemical pressure. 
At $y=0.7$ $\chi_{zz}$ is $T$-independent and remains diamagnetic.  
 
\begin{figure}
\includegraphics[width=9cm]{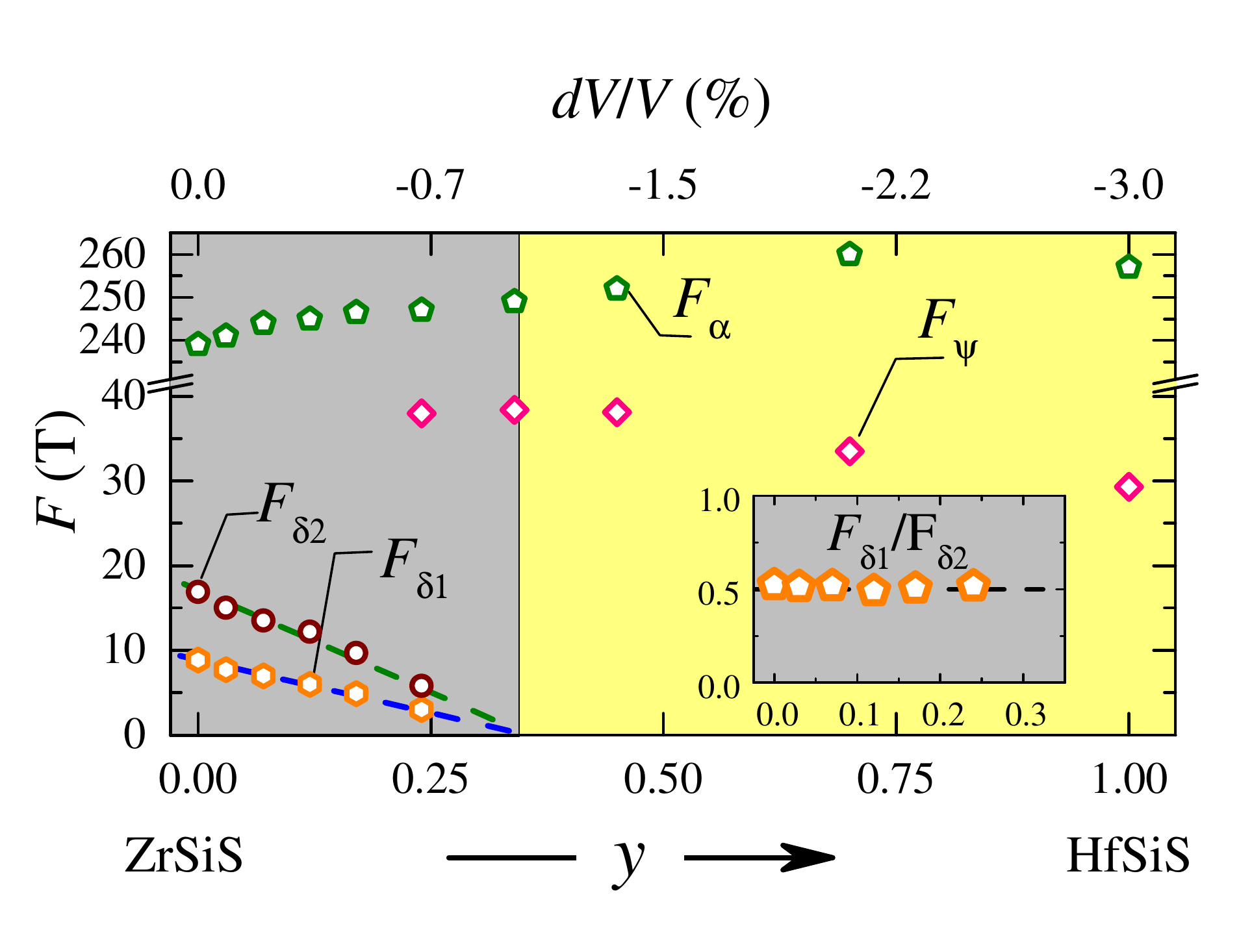}
\caption{\label{fig2} Phase diagram of  QO frequency vs Hf concentration  $y$  in Zr$_{1-y}$Hf$_y$SiS along the crystallographic  [001] direction.  Two Lifshitz transitions in the low-frequency spectrum are identified: A new pocket with frequency $F_{\psi}$ emerges at around $y=0.24$, and an electron pocket with frequency $F_{\delta1}$ ($F_{\delta2}$) vanishes at $y\approx0.34$ (as estimated from linear interpolations - green and blue dashed lines, respectively). 
Clearly, the low-frequency QOs seen in [001] direction for ZrSiS and HfSiS are  topologically not connected. 
On the other hand the hole-pocket frequency  $F_{\alpha}$ increases with $y$ and  is continuously connected between $y=0$ and $y=1$.  
The inset shows the ratio of the frequencies $F_{\delta1}/F_{\delta2}$ which  does not change with $y$.
The dashed lines are guide to the eye.}
\end{figure}

 To investigate whether  the unconventional behavior of $\chi_{zz}$ is linked to the change of the FS shape  we have performed QO measurements  for  $B||[001]$. 
 Figure \ref{fig2} shows  the QO frequencies as a function of chemical pressure. 
 It can be seen that the frequency  $F_{\alpha}$ associated with a hole-pocket  orbit in the $Z-R-A$  plane  smoothly increases with  $y$. 
 On the other hand, in the low-frequency part of the diagram two  Lifshitz transition can be identified. 
 The first one is an ordinary 2$\frac{1}{2}$ \cite{Lifshitz1960} Lifshitz transition  associated with the appearance of a new pocket with frequency $F_{\psi}$.
It  is detected at around $y=0.24$  and possibly could   be ascribed to  the appearance of an  electron pocket at the  $\Gamma$ point as predicted by  first-principle calculations in HfSiS \cite{Xu2015}.

 \begin{figure*}
\includegraphics[width=18cm]{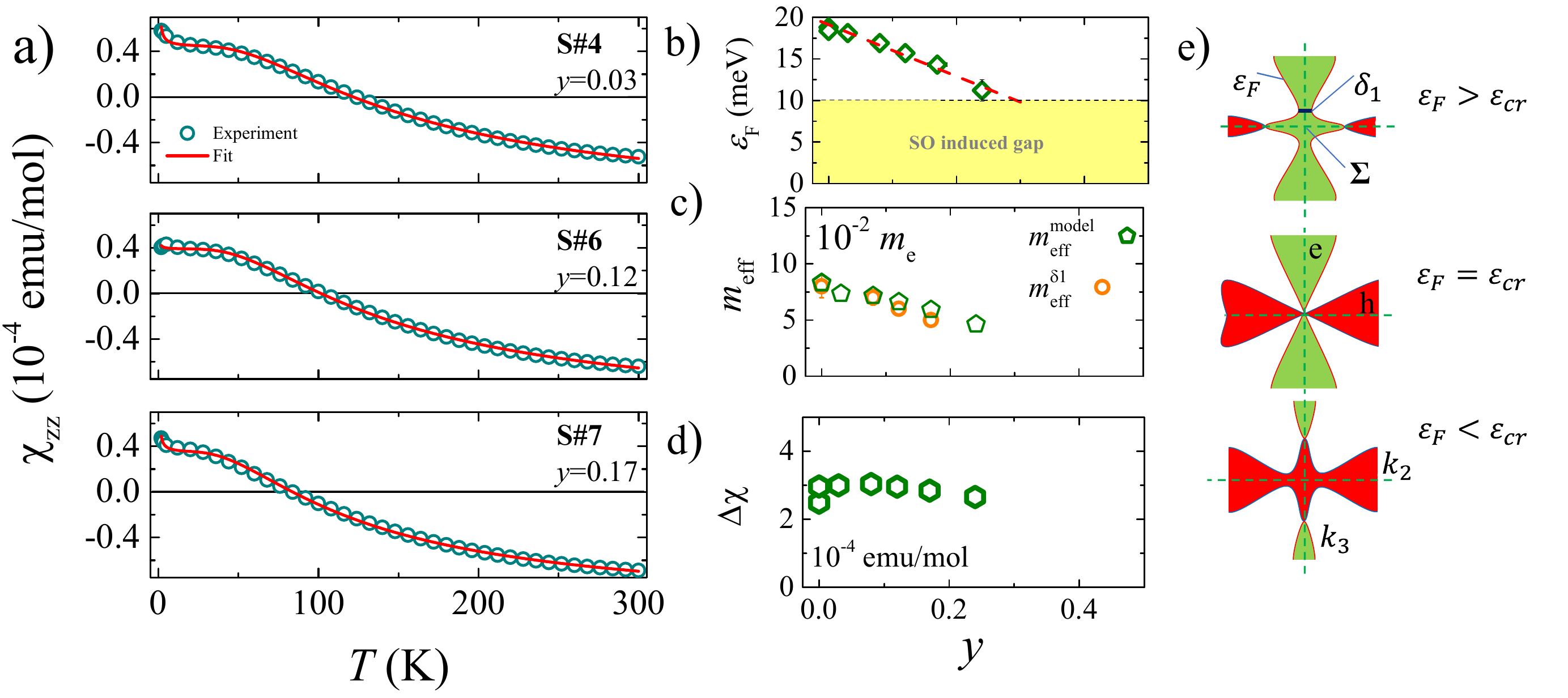}
\caption{\label{fig3} a) Fit of Eq.(\ref{2})  to experimental $\chi_{zz}(T)$  for samples $y=0.03$, $0.12$ and $0.17$ including localized  paramagnetic impurity contributions  responsible for the low-$T$ upturn. The model fairly well captures the experimental data in the whole measured $T$-range. 
b) Fermi energy ($E_{\rm F}$)  measured from   the nodal-line crossing point as a function of $y$. Below  $\approx 10$ meV our model breaks down due to the SO-interaction induced gap at the nodal line. 
c) Effective mass  $m^{\delta1}_{\rm eff}$ as estimated from QO of the $\delta1$ orbital compared with the effective mass $m^{model}_{\rm eff}$     as a function of $y$.
d) Almost constant value of  $\Delta \chi$ from Eq.(\ref{2}) as a function of $y$  indicates that the FS around the crossing points is not  strongly modified for low levels of $y$. 
e) Schematic  drawing of the Fermi surface with  self-intersecting orbitals near the  degeneracy point $\Sigma$  based on the model Eq.(\ref{1}) (parameters $B_1=-B_2$  and $\lambda<0$). By tuning the Fermi  energy the system  undergoes  a nonstandard  Lifshitz transition characterized by a change in the configuration of the  e-h pockets and direction of the  intersecting orbitals. }
\end{figure*}

The second  Lifshitz transition is associated with the continuously traced disappearance of  the oscillations with frequency $F_{\delta1}$  at around $y=0.3$. 
As shown in the inset to Fig.\ref{fig2},  the ratio of $F_{\delta1}$  and $F_{\delta2}$  is  constant and equal to 0.5. 
Thus we believe $F_{\delta2}$  originates from the  second harmonic response of the QOs and is not indicative of   an additional pocket.  
Interestingly, the disappearance of $F_{\delta1}$   coincides with  the disappearance of  the anomalous step-like behavior in the  susceptibility: 
For $y\leq 0.24$, the value of the  frequency $F_{\delta1}$ decreases with $y$.
At the  same time the sign change in $\chi_{zz}$ from diamagnetic to paramagnetic shifts to lower $T$ and for  $y=0.34$ $F_{\delta1}$ cannot be detected anymore, although   $\chi_{zz}(T)$ still exhibits a small sign change.
However, the latter   deviates considerably   from the step-like behavior observed for smaller $y$.


 $F_{\delta 1}$  seems to be produced by the FS cross section located near the $\Sigma$ point \cite{Muller2020, Mikitik2020}, a degeneracy point  at which the two nodal lines intersect  \cite{Schoop2016,Pezzini2018}, see Fig.~1c.
We can show that the observed temperature dependence of the magnetic susceptibility is also caused by the electron states in the vicinity of the $\Sigma$ point.
Neglecting the spin-orbit interaction, the dispersion of the two contacting bands ``$c$'' and ``$v$'' in the vicinity of this  crossing  point can be  approximately described by the following expression \cite{Mikitik2020,Mikitik2007}:
 \begin{eqnarray}\label{1}
 \varepsilon_{c,v}({\bf k})\approx B_2 k_2^2 + B_3 k_3^2\pm
 \left[(v_F k_1)^2+ \beta^2k_2^2k_3^2\right]^{1/2}\!\!,
 \end{eqnarray}
where $v_F$, $B_1$, $B_2$, $\beta$  are constant parameters of the spectrum, all the quasi-momenta $k_1$, $k_2$, $k_3$ are measured from the $\Sigma$ point, and $\varepsilon_{c,v}$ are measured from the energy of this point, $\varepsilon_{\Sigma}$. 
The $k_1$ axis coincides with the twofold symmetry axis $\Gamma$-M along which the crossing point $\Sigma$ is located.
The  $k_2$ and $k_3$ axes are along the tangents to the nodal lines crossing the $\Sigma$ point (in particular, the $k_3$ axis coincides with the $z$ axis); see Fig.~1. According to Ref.~\cite{Mikitik2007}, electron states in the vicinity of such a crossing point really can lead to an unusual anomaly in the magnetic susceptibility if the Fermi energy ($E_F$) is close to  $\varepsilon_{\Sigma}$. One has the following expression for the total magnetic susceptibility in this case \cite{Mikitik2021,Mikitik2007}:
 \begin{eqnarray}\label{2}
 \chi_{zz}=\chi_0+\Delta\chi[1+\exp(-E_F/k_BT)]^{-1},
  \end{eqnarray}
where $E_F$ is measured from the energy of the crossing point, and   $\chi_0$ is the magnetic susceptibility produced by the electron states lying far away from this point. 
The constant $\Delta\chi$ is defined by the  parameters of dispersion relation (\ref{1}), and at $|\lambda|\equiv 4|B_2B_3|/\beta^2\ll 1$, it has the form \cite{Mikitik2021, Mikitik2007}:
 \begin{eqnarray}\label{3}
\Delta\chi\approx -4\frac{e^2}{6\pi^2\hbar c^2}\frac{v_F\beta}{B_3},
 \end{eqnarray}
where we have taken into account that in the case of ZrSiS four $\Sigma$ points exist in the 1st Brillouin zone.

Formula (\ref{2}) with the addition of a $1/T$ term that is important below 10 K in order to account for the localized magnetic impurities reproduces the $T$-behavior of $\chi_{zz}$ extremely well in the low-$y$ region, Fig.~3a, if the chemical potential $E_F$ monotonically decreases with $y$, tending to the energy of the $\Sigma$ point, Fig.~3b. 
The parameter $\Delta \chi$ remains practically constant with changing $y$, Fig.~3d, which in the agreement with the natural assumption that  the parameters $v_F$, $\beta$, and $B_3$ are constant or  their variation is proportionally small  to the small $y$ \cite{interestingly}.

For $y>0.3$, formula (\ref{2}) ceases to fit the experimental data. This failure of Eq.~(\ref{2}) at $E_F\sim 10$ meV is due to the neglect of the spin-orbit interaction in deriving this formula. Note that the magneto-optical spectroscopy \cite{Uykur2019} shows that the gap in the spectrum induced by this interaction is of the order of $2\Delta \approx 26$ meV, and hence is  comparable with the limiting value of $E_F\sim 10$ meV. 
It is clear that at such a low value of $E_F$, one cannot neglect the spin-orbit gap.

Visible $y$ correlation between $F_{\delta 1}$ and the Fermi-level  $E_F$ extracted from the magnetic susceptibility indicate that the susceptibility and the frequency  $F_{\delta 1}$ are really  determined by the same part of the FS near the $\Sigma$ point.
To support this conclusion quantitatively, on Fig.\ref{fig3}c we compare experimental QO  effective masses with the one predicted by the model. 
The  model predicts the effective mass of  the cyclotron orbit around the neck of  the electron tube-like pocket near the $\Sigma$ to be $m^{\rm model}_{\rm eff}=3\hbar e F/2E_{\rm F}$, where $F$ is the frequency of QOs of the neck orbitals, and $E_F$ is the Fermi energy of the pocket \cite{Mikitik2020}.
Using the above-presented values yields   effective masses that are in excellent agreement with those  obtained from the QOs. 
The model can be further extended  by taking into account a spectrum with  the spin-orbit  gap \cite{privateconvo}, suggesting that the gap does not exceed  10 meV. 
For $y>0.3$, $E_F$ is likely located inside the spin-orbit gap at the $\Sigma$ point, and so the gap effects become important.
 Indeed, if we check the FS shape obtained by the first-principle calculations for HfSiS \cite{Kumar2017} and \footnote{See the Supplementary.}, we  see a FS with non-occupied states near the $\Sigma$ point. 

In conclusion,  we present a comprehensive study of the anomalous magnetic susceptibility of ZrSiS, which arises due to the proximity of the Fermi energy to the  degeneracy point formed by intersecting nodal lines.
The anomalous   susceptibility   provides a rare opportunity to observe   nodal-line physics by   thermodynamic methods. 
Introduction of   chemical pressure by replacing Zr with Hf in Zr$_{1-y}$Hf$_y$SiS  allows   to trace the evolution of the susceptibility anomaly and to test the model of the electronic spectrum around the $\Sigma$ degeneracy point. 
The model predicts  a temperature-dependent orbital magnetization which is in excellent agreement with the experimental data for Hf concentrations $y\leq0.24$. Also, the estimated spin-orbit gap at the nodal line matches the gap value obtained from the optical spectroscopy measurements. 
Moreover, the effective masses as estimated from  quantum oscillations are in excellent agreement with those yielded within our theoretical model.  
The phase diagram of Zr$_{1-y}$Hf$_y$SiS reveals two different   Lifshitz transitions,  indicating a  very rich  topology-related physics which could  also manifest in  other closely related nodal-line  systems \cite{Tremel1987,Xu2015}.

\textbf{Acknowledgements:} This work was supported by the CSF under the project IP 2018 01 8912 and CeNIKS project co-financed by the Croatian Government and the
EU through the European Regional Development Fund–Competitiveness and Cohesion Operational Program (Grant No. KK.01.1.1.02.0013). 
We thank Gaku Eguchi for the specific heat measurements and J.R. Cooper for giving constructive suggestions. 
A.K. acknowledges financial support from KAKENHI (Grants No.17H06138 and No.18H03683). M.N. acknowledges the financial support from JSPS through Invitational Fellowships for Research in Japan (Long-term).


\begin{thebibliography}{1}%
\makeatletter
\providecommand \@ifxundefined [1]{%
 \@ifx{#1\undefined}
}%
\providecommand \@ifnum [1]{%
 \ifnum #1\expandafter \@firstoftwo
 \else \expandafter \@secondoftwo
 \fi
}%
\providecommand \@ifx [1]{%
 \ifx #1\expandafter \@firstoftwo
 \else \expandafter \@secondoftwo
 \fi
}%
\providecommand \natexlab [1]{#1}%
\providecommand \enquote  [1]{``#1''}%
\providecommand \bibnamefont  [1]{#1}%
\providecommand \bibfnamefont [1]{#1}%
\providecommand \citenamefont [1]{#1}%
\providecommand \href@noop [0]{\@secondoftwo}%
\providecommand \href [0]{\begingroup \@sanitize@url \@href}%
\providecommand \@href[1]{\@@startlink{#1}\@@href}%
\providecommand \@@href[1]{\endgroup#1\@@endlink}%
\providecommand \@sanitize@url [0]{\catcode `\\12\catcode `\$12\catcode
  `\&12\catcode `\#12\catcode `\^12\catcode `\_12\catcode `\%12\relax}%
\providecommand \@@startlink[1]{}%
\providecommand \@@endlink[0]{}%
\providecommand \url  [0]{\begingroup\@sanitize@url \@url }%
\providecommand \@url [1]{\endgroup\@href {#1}{\urlprefix }}%
\providecommand \urlprefix  [0]{URL }%
\providecommand \Eprint [0]{\href }%
\providecommand \doibase [0]{https://doi.org/}%
\providecommand \selectlanguage [0]{\@gobble}%
\providecommand \bibinfo  [0]{\@secondoftwo}%
\providecommand \bibfield  [0]{\@secondoftwo}%
\providecommand \translation [1]{[#1]}%
\providecommand \BibitemOpen [0]{}%
\providecommand \bibitemStop [0]{}%
\providecommand \bibitemNoStop [0]{.\EOS\space}%
\providecommand \EOS [0]{\spacefactor3000\relax}%
\providecommand \BibitemShut  [1]{\csname bibitem#1\endcsname}%
\let\auto@bib@innerbib\@empty
\bibitem [{Note1()}]{Note1}%
  \BibitemOpen
  \bibinfo {note} {See the Supplementary.}\BibitemShut {Stop}%
\end{thebibliography}%


\begin{thebibliography}{9}
\bibitem{Geim2007} A. K. Geim and K. S. Novoselov, The rise of graphene, Nature Materials \textbf{6}, 183 (2007).
\bibitem{Armitage2018} N. P. Armitage, E. J. Mele, and A. Vishwanath, Weyl and
Dirac semimetals in three-dimensional solids, Reviews of
Modern Physics \textbf{90}, 015001 (2018), arXiv:1705.01111.
\bibitem{Xiong2015}  J. Xiong, S. K. Kushwaha, T. Liang, J. W. Krizan,
M. Hirschberger, W. Wang, R. J. Cava, and N. P. Ong,
Evidence for the chiral anomaly in the Dirac semimetal
Na3Bi, Science \textbf{350}, 413 (2015).
\bibitem{McClure1956}
J. W. McClure, Diamagnetism of Graphite, Physical Review \textbf{104}, 666 (1956).
\bibitem{Plochocka2008}
P. Plochocka, C. Faugeras, M. Orlita, M. L. Sadowski,
G. Martinez, M. Potemski, M. O. Goerbig, J. N. Fuchs,
C. Berger, and W. A. De Heer, High-energy limit of massless dirac fermions in multilayer graphene using magnetooptical transmission spectroscopy, Physical Review Letters \textbf{100}, 1 (2008), arXiv:0709.1324.
\bibitem{Martino2019} E. Martino, I. Crassee, G. Eguchi, D. Santos-Cottin,
R. D. Zhong, G. D. Gu, H. Berger, Z. Rukelj, M. Orlita, C. C. Homes, and A. Akrap, Two-Dimensional Conical Dispersion in ZrTe5 Evidenced by Optical Spectroscopy, Physical Review Letters \textbf{122}, 217402 (2019),
arXiv:1905.00280.
\bibitem{Mikitik1989}G. Mikitik and I. Svechkarev, Giant Anomalies of
Magnetic-Susceptibility Due To Energy-Band Degeneracy in Crystals, Sov. J. Low Temp. Phys. \textbf{15}, 165 (1989).
\bibitem{Mikitik2016}G. P. Mikitik and Y. V. Sharlai, Magnetic susceptibility
of topological nodal semimetals, Physical Review B \textbf{94},
195123 (2016).

\bibitem{Koshino2007} M. Koshino and T. Ando, Diamagnetism in disordered
graphene, Physical Review B \textbf{75}, 235333 (2007).
\bibitem{Koshino2016} M. Koshino and I. F. Hizbullah, Magnetic susceptibility
in three-dimensional nodal semimetals, Physical Review
B \textbf{93}, 045201 (2016).
\bibitem{Ominato2018}Y. Ominato and K. Nomura, Spin susceptibility of threedimensional Dirac-Weyl semimetals, Physical Review B
\textbf{97}, 245207 (2018).
\bibitem{Ogata2015}M. Ogata and H. Fukuyama, Orbital Magnetism of Bloch
Electrons I. General Formula, Journal of the Physical
Society of Japan \textbf{84}, 124708 (2015), arXiv:1602.02449.
\bibitem{Fukuyama1971}H. Fukuyama, Theory of Orbital Magnetism of Bloch
Electrons: Coulomb Interactions, Progress of Theoretical Physics \textbf{45}, 704 (1971).
\bibitem{Raoux2014}A. Raoux, M. Morigi, J.-N. Fuchs, F. Piéchon, and
G. Montambaux, From Dia- to Paramagnetic Orbital
Susceptibility of Massless Fermions, Physical Review Letters \textbf{112}, 026402 (2014).
\bibitem{Nair2018}N. L. Nair, P. T. Dumitrescu, S. Channa, S. M. Griffin,
J. B. Neaton, A. C. Potter, and J. G. Analytis, Thermodynamic signature of Dirac electrons across a possible
topological transition in ZrTe5, Physical Review B \textbf{97},
041111 (2018).
\bibitem{Fuseya2015}Y. Fuseya, M. Ogata, and H. Fukuyama, Transport properties and diamagnetism of dirac electrons in bismuth,
Journal of the Physical Society of Japan \textbf{84}, 54 (2015),
arXiv:1407.2179.
\bibitem{Tremel1987}W. Tremel and R. Hoffmann, Square nets of main-group
elements in solid-state materials, Journal of the American
Chemical Society \textbf{109}, 124 (1987).
\bibitem{Xu2015} Q. Xu, Z. Song, S. Nie, H. Weng, Z. Fang, and X. Dai,
Two-dimensional oxide topological insulator with ironpnictide superconductor LiFeAs structure, Physical Review B \textbf{92}, 205310 (2015).
\bibitem{Schoop2016}L. M. Schoop, M. N. Ali, C. Straßer, A. Topp,
A. Varykhalov, D. Marchenko, V. Duppel, S. S. P.
Parkin, B. V. Lotsch, and C. R. Ast, Dirac cone protected
by non-symmorphic symmetry and three-dimensional
Dirac line node in ZrSiS, Nature Communications \textbf{7},
11696 (2016).
\bibitem{Neupane2016}M. Neupane, I. Belopolski, M. M. Hosen, D. S.
Sanchez, R. Sankar, M. Szlawska, S.-Y. Xu, K. Dimitri, N. Dhakal, P. Maldonado, P. M. Oppeneer, D. Kaczorowski, F. Chou, M. Z. Hasan, and T. Durakiewicz,
Observation of topological nodal fermion semimetal
phase in ZrSiS, Physical Review B \textbf{93}, 201104 (2016),
arXiv:1604.00720.
\bibitem{Hirayama2017} M. Hirayama, R. Okugawa, T. Miyake, and S. Murakami,
Topological Dirac nodal lines and surface charges in fcc
alkaline earth metals, Nature Communications \textbf{8}, 14022
(2017), arXiv:1602.06501.
\bibitem{Pezzini2018}S. Pezzini, M. R. van Delft, L. M. Schoop, B. V. Lotsch,
A. Carrington, M. I. Katsnelson, N. E. Hussey, and S. Wiedmann, Unconventional mass enhancement around the Dirac nodal loop in ZrSiS, Nature Physics \textbf{14}, 178
\bibitem{Shao2020}Y. Shao, A. N. Rudenko, J. Hu, Z. Sun, Y. Zhu, S. Moon,
A. J. Millis, S. Yuan, A. I. Lichtenstein, D. Smirnov,
Z. Q. Mao, M. I. Katsnelson, and D. N. Basov, Electronic
correlations in nodal-line semimetals, Nature Physics \textbf{16},
636 (2020).
\bibitem{Gatti2020}G. Gatti, A. Crepaldi, M. Puppin, N. Tancogne-Dejean,
L. Xian, U. De Giovannini, S. Roth, S. Polishchuk,
P. Bugnon, A. Magrez, H. Berger, F. Frassetto, L. Poletto, L. Moreschini, S. Moser, A. Bostwick, E. Rotenberg, A. Rubio, M. Chergui, and M. Grioni, LightInduced Renormalization of the Dirac Quasiparticles in
the Nodal-Line Semimetal ZrSiSe, Physical Review Letters \textbf{125}, 076401 (2020), arXiv:1912.09673.
\bibitem{Fu2019}B.-B. Fu, C.-J. Yi, T.-T. Zhang, M. Caputo, J.-Z. Ma,
X. Gao, B. Q. Lv, L.-Y. Kong, Y.-B. Huang, P. Richard,
M. Shi, V. N. Strocov, C. Fang, H.-M. Weng, Y.-G. Shi,
T. Qian, and H. Ding, Dirac nodal surfaces and nodal
lines in ZrSiS, Science Advances \textbf{5}, eaau6459 (2019).

\bibitem{Orbanic2021}F. Orbanić, M. Novak, Z. Glumac, A. McCollam,
L. Tang, and I. Kokanović, Quantum oscillations of the
magnetic torque in the nodal-line dirac semimetal zrsis,
Phys. Rev. B \textbf{103}, 045122 (2021).
\bibitem{Uykur2019}E. Uykur, L. Z. Maulana, L. M. Schoop, B. V. Lotsch,
M. Dressel, and A. V. Pronin, Magneto-optical probe of
the fully gapped Dirac band in ZrSiS, Physical Review
Research \textbf{1}, 3 (2019), arXiv:1910.07571.
\bibitem{Muller2020}C. S. A. Müller, T. Khouri, M. R. van Delft, S. Pezzini,
Y.-T. Hsu, J. Ayres, M. Breitkreiz, L. M. Schoop, A. Carrington, N. E. Hussey, and S. Wiedmann, Determination
of the Fermi surface and field-induced quasiparticle tunneling around the Dirac nodal loop in ZrSiS, Physical
Review Research \textbf{2}, 023217 (2020).
\bibitem{Novak2019}M. Novak, S. N. Zhang, F. Orbanić, N. Biliškov,
G. Eguchi, S. Paschen, A. Kimura, X. X. Wang, T. Osada, K. Uchida, M. Sato, Q. S. Wu, O. V. Yazyev, and
I. Kokanović, Highly anisotropic interlayer magnetoresitance in ZrSiS nodal-line dirac semimetal, Phys. Rev. B
\textbf{100}, 085137 (2019).
\bibitem{VanGennep2019}D. VanGennep, T. A. Paul, C. W. Yerger, S. T. Weir,
Y. K. Vohra, and J. J. Hamlin, Possible pressure-induced
topological quantum phase transition in the nodal line
semimetal ZrSiS, Physical Review B \textbf{99}, 085204 (2019),
arXiv:1901.07043.
\bibitem{Gu2019}C. C. Gu, J. Hu, X. L. Chen, Z. P. Guo, B. T. Fu, Y. H.
Zhou, C. An, Y. Zhou, R. R. Zhang, C. Y. Xi, Q. Y. Gu,
C. Park, H. Y. Shu, W. G. Yang, L. Pi, Y. H. Zhang,
Y. G. Yao, Z. R. Yang, J. H. Zhou, J. Sun, Z. Q. Mao, and
M. L. Tian, Experimental evidence of crystal symmetry
protection for the topological nodal line semimetal state
in ZrSiS, Physical Review B \textbf{100}, 205124 (2019).
\bibitem{Lifshitz1960}I. M. Lifshitz, Anomalies of Electron Characteristics of
a Metal in the High Pressure Region, Sov. Phys. JETP
\textbf{11}, 1130 (1960).

\bibitem{Mikitik2020} G. P. Mikitik and Y. V. Sharlai, Crossing points of nodal lines in topological semimetals and the Fermi surface of ZrSiS, Physical Review B \textbf{101}, 205111 (2020),
arXiv:2003.00552.

\bibitem{Mikitik2007} G. P. Mikitik, Step-like anomaly of the magnetic susceptibility in crystals with degenerate electronic energy bands, Low Temperature Physics \textbf{33}, 839 (2007).

\bibitem{Mikitik2021}G. P. Mikitik and Y. V. Sharlai, Magnetic susceptibility of crystals with crossing of their band-contact lines, Low Temperature Physics \textbf{47}, 605 (2021).

\bibitem{interestingly}Interestingly, the values of these parameters ($v_F\approx 7.7\times 10^5$ m/s, $B_3\approx -0.25/m$, $\beta \approx 1.3/m$ where $m$ is the electron mass) roughly estimated in Ref.~\cite{Mikitik2020} give $\Delta \chi\approx 6.6\times 10^{-6}$, which is of the order of $\Delta\chi\approx 9.5\times 10^{-6}$ ($3\times 10^{-4}$emu/mol)  presented in Fig.~3d.

\bibitem{privateconvo}N. Novak, Yuriy V. Sharlai and Grigorii P. Mikitik  private communication.

\bibitem{Kumar2017}N. Kumar, K. Manna, Y. Qi, S.-c. Wu, L. Wang, B. Yan,
C. Felser, and C. Shekhar, Unusual magnetotransport
from Si-square nets in topological semimetal HfSiS, Physical Review B \textbf{95}, 121109 (2017).


\end{thebibliography}
\end{document}